\documentclass[aps,pra,twocolumn,superscriptaddress,amsmath,amssymb,letterpaper]{revtex4-1} 

\ifx\pdfoutput\@undefined\usepackage[usenames,dvips]{color}
\else\usepackage[usenames,dvipsnames]{color}

\usepackage[english]{babel}
\usepackage{amsfonts, amsmath,amssymb}
\usepackage{bm}
\usepackage{verbatim} 
\usepackage{graphics,graphicx}
\usepackage{epsfig}
\usepackage{epstopdf}
\usepackage{color,xcolor}
\definecolor{dark-green}{rgb}{0,0.625,0}
\definecolor{NEW}{rgb}{1,0,0}
\definecolor{NEW2}{rgb}{1,0,1}
\usepackage{hyperref}
\usepackage{natbib}
\usepackage{cleveref}				
\def\dfrac{\displaystyle\frac}  

\begin{document}
       
\title{Relativistic spin-orbit interactions of photons and electrons}
 
\author{D. A. Smirnova}
\thanks{These authors contributed equally to this work.}
\affiliation{Nonlinear Physics Centre, RSPE, The Australian National University,
Canberra, ACT 0200, Australia}

\author{V. M. Travin}
\thanks{These authors contributed equally to this work.}
\affiliation{V. N. Karazin Kharkov National University, 4 Svobody Sq., Kharkov 61022, Ukraine}
\affiliation{Institute for Low Temperature and Structure Research, Polish Academy of Sciences, POB 1410, 50-950 Wroclaw, Poland}

\author{K. Y. Bliokh}
\thanks{These authors contributed equally to this work.}
\affiliation{Nonlinear Physics Centre, RSPE, The Australian National University,
Canberra, ACT 0200, Australia}
\affiliation{CEMS, RIKEN, Wako-shi, Saitama 351-0198, Japan}

\author{F. Nori}
\affiliation{CEMS, RIKEN, Wako-shi, Saitama 351-0198, Japan}
\affiliation{Physics Department, University of Michigan, Ann Arbor, Michigan 48109-1040, USA}

\begin{abstract}
Laboratory optics, typically dealing with monochromatic light beams in a single reference frame, exhibits numerous spin-orbit interaction phenomena due to the coupling between the spin and orbital degrees of freedom of light. 
Similar phenomena appear for electrons and other spinning particles. 
Here we examine transformations of paraxial photon and relativistic-electron states carrying the spin and orbital angular momenta (AM) under the Lorentz boosts between different reference frames. We show that transverse boosts inevitably produce a rather nontrivial conversion from spin to orbital AM. The converted part is then separated between the intrinsic (vortex) and extrinsic (transverse shift or Hall effect) contributions. Although the spin, intrinsic-orbital, and extrinsic-orbital parts all point in different directions, such complex behavior is necessary for the proper Lorentz transformation of the total AM of the particle. Relativistic spin-orbit interactions can be important in scattering processes involving photons, electrons, and other relativistic spinning particles, as well as when studying light emitted by fast-moving bodies.
\end{abstract}

\maketitle 

\section{Introduction}

In the past decade, the {\it spin-orbit interactions} (SOIs) of light --- including spin Hall effects, spin-to-orbital angular momentum (AM) conversions, etc. --- have become an inherent part of modern optics, see \cite{Bliokh2015,Hasman2005,Bliokh2009,Marrucci2011,Xiao2016} for reviews. The vast majority of known SOI effects originate from the fundamental polarization and AM properties of monochromatic Maxwell fields in a {\it single laboratory reference frame} \cite{Bliokh2015,Bliokh2010}. 
Similar phenomena have also been described for relativistic electrons and other spinning particles \cite{Berard2006,Bliokh2005,Duval2006,Chang2008,Xiao2010,Bliokh2011}.
Electron SOIs play an important role in atomic physics, condensed matter, and could also affect the dynamics of relatvistic free-electron states carrying intrinsic AM \cite{Harris2015,Bliokh2017review,Lloyd2017}.

At the same time, there is considerable recent interest on {\it relativistic transformations} of photons and other particles carrying intrinsic AM \cite{Bliokh2012,Chen2014,Duval2015,Stone2015,Maslanka2017}. Transverse Lorentz boosts of wave beams break down their monochromaticity and induce a number of nontrivial relativistic AM-dependent phenomena. In particular, the Lorentz transformations of the intrinsic and extrinsic AM differ significantly from each other. Requiring their consistency brings about the relativistic Hall effect (i.e., the boost-induced transverse position shift) related to the delocalized nature of the wave AM \cite{Bliokh2012,Chen2014,Duval2015,Stone2015,Maslanka2017}. 

Relativistic properties and transformations of AM-carrying waves are important from both the fundamental and practical viewpoints. These are involved in the ``proton spin puzzle'' in QCD \cite{Leader,Wakamatsu}, studies of ``chiral fermions'' \cite{Chen2014,Duval2015}, and collisions of spinning particles \cite{Chen2014,Stone2015}. Moreover, there is a rapidly growing interest in scattering of photons, electrons, and other high-energy particles carrying intrinsic orbital AM \cite{Bliokh2017review,Serbo2011,Ivanov2011,Karlovets2017}. Naturally, the Lorentz transformations of wavepackets or beams carrying spin and orbital AM are of great importance for these topics.

Importantly, most of the recent studies of Lorentz transformations of the wave AM considered the intrinsic spin (polarization) and orbital (vortex) AM on equal footing. For example, the Hall-effect shift of the energy centroid is largely independent of the spin or orbital nature of the intrinsic AM \cite{Bliokh2012,Chen2014,Stone2015,Maslanka2017,Muller1992}. 
However, in this paper, we show that the spin and orbital AM of photons and relativistic electrons are transformed quite differently under Lorentz boosts. To illustrate this crucial difference, we put forward the following paradox about the transformation of the spin of a photon. 

\begin{figure}[b]
\centering
\includegraphics[width=\columnwidth]{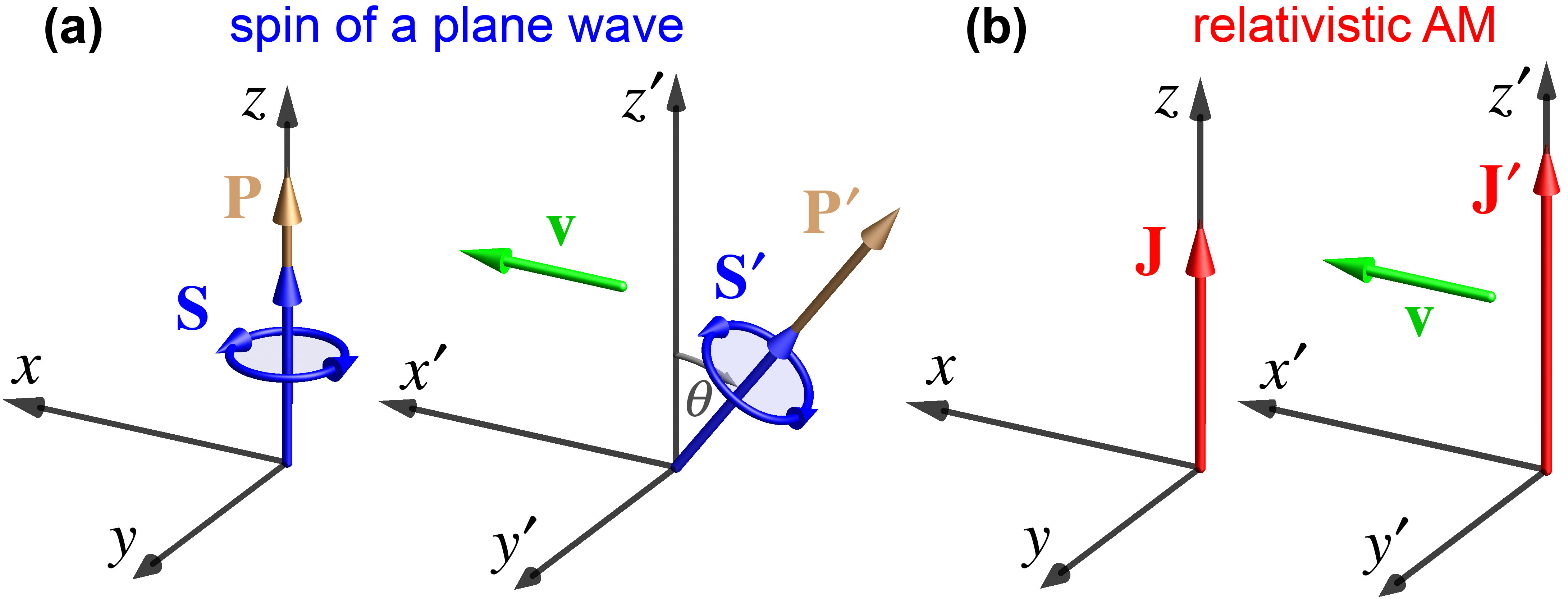}
\caption{Transverse Lorentz transformation of the photon spin. The spin of an electromagnetic wave, ${\bf S}$, is rotated by the angle $\theta = \sin^{-1}\!u$ (a), which is in contrast to the Lorentz transformation of a relativistic AM ${\bf J}$ (provided the boost momentum ${\bf N}={\bf 0}$): ${\bf J}^\prime = \gamma {\bf J}$ (b) \cite{LL,Bliokh2012}.}
\label{Fig_1}
\end{figure}

The spin AM of a paraxial photon can be well approximated by the plane-wave expression ${\bf S} = \hbar \sigma {\bf P}/P$ \cite{OAM,AML,Bliokh2015PR}, where $\sigma \in (-1,1)$ is the polarization helicity and ${\bf P}$ is the photon momentum. Assuming ${\bf P} = P\, \bar{\bf z}$ (overbars denote the unit vectors of the corresponding axes), we perform the transverse Lorentz boost characterized by the velocity ${\bf v} = v\, \bar{\bf x}$ and the corresponding Lorentz factor $\gamma = 1/\sqrt{1-u^2}$, $u\equiv v/c$. This transformation preserves the helicity $\sigma$ (which is Lorentz-invariant for massless particles) and rotates the propagation direction by the angle $\theta = \sin^{-1}\!u = \cos^{-1}\!\gamma^{-1}$. As a result, the photon spin in the boosted reference frame (indicated by primes) becomes ${\bf S}^\prime = \hbar\sigma[\gamma^{-1}\bar{\bf z}^\prime - u\, \bar{\bf x}^\prime]$, Fig.~1(a). However, this contradicts the Lorentz transformations of the relativistic AM tensor, which consists of the the AM ${\bf J}$ and the ``boost momentum'' ${\bf N}$ \cite{LL}. Indeed, assuming that the photon is represented by a large paraxial wavepacket (close enough to the plane wave) with energy $W=\hbar\omega$ ($\omega=kc$ is the frequency, ${\bf k}$ is the wavevector), momentum ${\bf P} = \hbar k\, \bar{\bf z}$, and position of the centroid ${\bf R} = ct\, \bar{\bf z}$, the boost momentum vanishes in the original reference frame: ${\bf N} = ct {\bf P} -{\bf R}W/c = {\bf 0}$. Then, the Lorentz transformations of the AM ${\bf J} = J\, \bar{\bf z}$ yields ${\bf J}^\prime = \gamma\, {\bf J}$ \cite{LL,Bliokh2012,remark}, Fig.~1(b). Obviously, the photon spin AM ${\bf S}$ cannot follow this rule because $\gamma > 1$, while the spin is restricted to the $(-\hbar,\hbar)$ range. 
In a more fundamental context, the difference between the spin and AM transformations come from the fact that the spin of a relativistic particle follows the Pauli-Lubanski four-vector rather than the rank-2 AM tensor \cite{QED,Ivan,Bliokh2017}.

In this paper, we resolve the above controversy by considering trasversely-localized optical beams carrying both spin and orbital AM. We derive quite nontrivial Lorentz transformations of the spin and orbital AM carryied by paraxial photons, as well as the relativistic Hall-effect shifts caused by photon's spin and orbital AM. 
We find that Lorentz boosts inevitably produce spin-to-orbital AM conversion as well as nontrivial spin and orbital Hall-effect shifts, i.e., {\it relativistic SOIs}. We also perform analogous Lorentz-boost calculations for the Dirac-electron beams, and show that most of their AM transformation features are similar to the photon case, albeit modified by the finite electron mass.

\section{Relativistic transformations of optical beams}
\subsection{General formalism}

We first introduce the general formalism for calculations of dynamical properties (energy, momentum, AM, etc.) of generic free-space Maxwell fields. This is mosty based on the results of works \cite{Bliokh2010,BBreview,BB2011}.

The real electric and magnetic fields $\bm{\mathcal{E}}(t, {\bf r})$ and $\bm{\mathcal{H}}
(t, {\bf r})$ are represented via their complex Fourier (plane-wave) components:
%
\begin{eqnarray}
\label{eq:Fourier}
\left\{ {\begin{array}{c}
   {\bm {\mathcal{E}}}(t, {\bf r})  \\
   {\bm {\mathcal{H}}}(t, {\bf r})  
\end{array}} \right\}  =\! 2 {\rm Re}\! \int\!{\frac{d^3{\bf k}}{(2\pi)^{3/2}}} 
\left\{ 
{\begin{array}{c}
   {\bf E}({\bf k})  \\
{\bf H}({\bf k})
\end{array}}
\right\}
e^{-i\omega t+i{\bf k}\cdot{\bf r} },
\end{eqnarray}
%
where $\omega({\bf k}) = kc$. Due to Maxwell's equations, the Fourier components are orthogonal to the wavevector: ${\bf E} \cdot {\bf k} = {\bf H} \cdot {\bf k} =0$, and it is instructive to make a transformation to the local ${\bf k}$-space coordinates with the longitudinal axis attached to the wavevector. The fields have only two transverse components in these coordinates, and using the basis of circular polarizations corresponds to the {\it helicity representation} of Maxwell fields. 

The transition to this basis is realized by the unitary transformation \cite{Bliokh2010} 
%
\begin{eqnarray} 
\label{eq:htransform}
& \left\{ \tilde{{\bf E}} ({\bf k}), \tilde{{\bf H}} ({\bf k}) \right\} = \hat{V}\, \hat{U}({\bf k}) \left\{ {\bf E} ({\bf k}), {\bf H} ({\bf k}) \right\}~.
\end{eqnarray} 
%
Here 
%
\begin{eqnarray} 
& \hat{U}({\bf k}) = \hat{\mathcal R}_z(-\varphi) \hat{\mathcal R}_y(\vartheta) \hat{\mathcal R}_z(\varphi) \nonumber \\
& = \! \left( \begin{array}{ccc}
1 \! - \! 2\sin^2\!\frac{\vartheta}{2}\cos^2\!\varphi & -\sin^2\!\frac{\vartheta}{2}\sin2\varphi & -\sin\vartheta\cos\varphi  \\
-\sin^2\!\frac{\vartheta}{2}\sin2\varphi  & 1 \! - \! 2\sin^2\!\frac{\vartheta}{2}\sin^2\!\varphi & -\sin\vartheta\cos\varphi \\
\sin\vartheta\cos\varphi & \sin\vartheta\sin\varphi  & \cos\vartheta \end{array} \right) \nonumber
\end{eqnarray} 
%
is the rotational matrix superimposing the longitudinal axis with the wavevector ($(\vartheta,\varphi)$ are the spherical angles of the ${\bf k}$-vector and $\hat{\mathcal R}_{y,z}$ are the corresponding rotational matrices), 
whereas 
%
\begin{eqnarray} 
\hat{V}= \dfrac{1}{\sqrt{2}}\left( \begin{array}{ccc}
1 & -i & 0 \\
1 & i & 0 \\
0 & 0 & {\sqrt{2}} \end{array} \right). \nonumber 
\end{eqnarray} 
%
is the constant matrix of the transition to the circular-polarization basis.

Omitting the vanishing longitudinal component of the fields (\ref{eq:htransform}), we end up with the {\it two-component} electric field $\tilde{{\bf E}} = \left( \tilde{E}^+ , \tilde{E}^- \right)^T$ and the corresponding magnetic field $\tilde{\bf H} = -i \hat{\sigma} \tilde{\bf E}$ following from Maxwell's equations. Here $\hat{\sigma} = {\rm diag}(1,-1)$ is the {\it helicity operator} and throughout the paper we use Gaussian-like units with $\varepsilon_0=\mu_0=1$.

We now define the ``{\it photon wavefunction}'' \cite{BBreview,BB2011} in the helicity representation as 
%
\begin{equation} 
\label{eq:wf}
 \bm{\psi}({\bf k}) = \frac{1}{\sqrt{2\mathcal N}} \left[ \tilde{\bf E}({\bf k}) + i \hat{\sigma} \tilde{\bf H} ({\bf k}) \right] =  \sqrt{\frac{2}{\mathcal N}}\, \tilde{\bf E}({\bf k})~,
\end{equation} 
%
where the normalization factor ${\mathcal N}$ is the number of photons defined below.
Then, the expectation value of an operator $\hat{O}$ can be calculated as \cite{BBreview,BB2011}
%
\begin{equation} 
\label{eq:EV}
O = \langle{\bm{\psi}}|{\hat O}|{\bm{\psi}}\rangle \equiv \int\frac{d^3{\bf k}}{\hbar\omega} \, \bm{\psi}^\dagger({\bf k})\! \cdot \! (\hat{O})\, \bm{\psi}({\bf k})~.
\end{equation} 
%
Note that the factor $\omega^{-1}({\bf k})$ in Eq.~(\ref{eq:EV}) is crucial for non-monochromatic fields. 
Assuming the one-photon normalization $\langle{\bm{\psi}}|{\bm{\psi}}\rangle =1$, the number of photons in Eq.~(\ref{eq:wf}) is ${\mathcal N} = 2 \int \frac{d^3{\bf k}}{\hbar\omega} \, | \tilde{\bf E}({\bf k})|^2$.

For further calculations, we need operators of the energy $W$, momentum ${\bf P}$, position ${\bf R}$, spin AM ${\bf S}$, orbital AM ${\bf L}$, and boost momentum ${\bf N}$ (see \cite{LL,Bliokh2012,Barnett2011,Bliokh2013,Cameron2012} for the latter quantity). According to the works \cite{Bliokh2010,BBreview,BB2011}, in the helicity representation, 
projected on the 2D subspace of Maxwell fields $\tilde{\bf E}({\bf k})$, 
these operators read
%
\begin{eqnarray}
{\hat{W}}=\hbar\omega, & \quad & {\hat{\bf P}}=\hbar{\bf k},  \quad {\hat{\bf R}}= i  {\bm{\nabla}}_{{\bf k}} - {\hat{\bf A}}_{B} ({\bf k}), \nonumber \\
{\hat{\bf S}}= \hbar {\hat \sigma } \frac{\bf k}{k} , & \quad &
{\hat{\bf L}}= {\hat{{\bf R}}} \times {\hat{\bf P}},  \quad  {\hat{\bf N}} = ct \hat{\bf P} - \hat{\bf R}\hat{W}/c .
\label{eq:operators}
\end{eqnarray}  
%
Here ${\hat{\bf A}}_{B} ({\bf k}) = \hat{\sigma} k^{-1}\left[(1-\cos\vartheta)/\sin\vartheta\right]\, \bar{\bm \varphi} $ ($\bar{\bm \varphi}$ is the unit vector of the azimuthal coordinate $\varphi$) is the {\it Berry connection}, which determines the covariant derivative and parallel transport of ${\bf E}({\bf k})\perp {\bf k}$ on the sphere $S^2 = \{ {\bf k}/k \}$ 
(the electric field ${\bf E}({\bf k})$ belongs to the vector fiber bundle over this sphere) 
\cite{Bliokh2015,Bliokh2009,Bliokh2010,BB2011,BB1987}. 
The operators $\hat{W}$, $\hat{\bf P}$, the total AM $\hat{\bf J} = \hat{\bf S} +\hat{\bf L}$, and $\hat{\bf N}$ provide 10 generators of the {\it Poincar\'{e} group}, and their expectation values are conserved in free space \cite{BBreview,BB2011,Bliokh2013,Cameron2012}.

Note that the expectation value of the position operator (\ref{eq:operators}), ${\bf R}$, describes the ``{\it photon centroid}'', while the {\it energy centroid} of the field is determined as ${\bf R}_E = c(ct{\bf P}-{\bf N})/W$. These two positions can differ from each other in non-monochromatic fields; they play a crucial role in the Lorentz transformations of the AM and relativistic Hall effects \cite{LL,Bliokh2012,Chen2014,Stone2015,Muller1992}. 

We also note that the same expectation values (\ref{eq:EV}) can be obtained without transition to the helicity representation (\ref{eq:htransform}). In the canonical momentum representation, the 6-component ``photon wavefunction'' is given by the Fourier components (\ref{eq:Fourier}):
%
\begin{equation}
\bm{\psi}_{\rm can}(\bf{k}) = \frac{1}{\sqrt{\mathcal N}}
\left\{ {\bf E}({\bf k}), {\bf H}({\bf k})\right\}~,
\label{eq:wfcan}
\end{equation}  
%
where ${\mathcal N} = \int \frac{d^3{\bf k}}{\hbar\omega} \left( | {\bf E}({\bf k})|^2+ | {\bf H}({\bf k})|^2 \right)$. In this representation, the operators (\ref{eq:operators}) have canonical form without the Berry connection:
%
\begin{eqnarray}
{\hat{\bf R}_{\rm can}} & = & i  {\bm{\nabla}}_{{\bf k}}, \quad
{\hat{\bf L}}_{\rm can}= {\hat{{\bf R}}_{\rm can}} \times {\hat{\bf P}},  \nonumber \\  
{\hat{\bf N}}_{\rm can} & = & ct \hat{\bf P} - \hat{\bf R}_{\rm can}\hat{W}/c ,
\label{eq:operatorscan}
\end{eqnarray}  
%
whereas the spin operator ${\hat{\bf S}}_{\rm can}$ is given by the momentum-independent spin-1 $3 \times 3$ matrices \cite{Bliokh2010,BBreview,Bliokh2015PR}.
Although the canonical operators have simpler form, the canonical photon wavefunction (\ref{eq:wfcan}) is considerably complicated, having six components instead of two. Therefore, below we employ the helicity representation (\ref{eq:htransform})--(\ref{eq:operators}) for photonic calculations, but use the canonical representation analogous to Eqs.~(\ref{eq:operatorscan}) for the Dirac-electron calculations in Section III.

In addition to the expectation values of operators (\ref{eq:operators}) in the momentum representation, we will use the spatial energy and Poynting-momentum densities in the coordinate representation of real fields (\ref{eq:Fourier}) \cite{LL}:
%
\begin{equation}
w =\left( {\bm {\mathcal{E}}}^2 + {\bm {\mathcal{H}}}^2 \right)/2,
\quad
{\bf p} = {c}^{-1}\! \left( {\bm {\mathcal{E}}} \times {\bm {\mathcal{H}}} \right).
\label{eq:densities}
\end{equation}  
%
The integral energy, momentum, total AM, and boost momentum of a localized field are then determined as $W=\int w\, d^3{\bf r}$, ${\bf P}=\int {\bf p}\, d^3{\bf r}$, ${\bf J}=\int ({\bf r}\times{\bf p})\, d^3{\bf r}$, and ${\bf N}=\int (ct{\bf p} - {\bf r} w/c)\, d^3{\bf r}$. For a one-photon field, these values are equivalent to the corresponding expectation values calculated using Eqs.~(\ref{eq:wf})--(\ref{eq:operators}).

We finally describe the Lorentz boosts of a generic electromagnetic field. The real fields $\left\{ {\bm {\mathcal{E}}}(t, {\bf r}),{\bm {\mathcal{H}}}(t, {\bf r}) \right\}$ are transformed as components of the antisymmetric rank-2 field tensor, together with the Lorentz transformation of the four-coordinates $(ct,{\bf r})$ \cite{LL}. 
The Fourier components  $\left\{ {\bf E}({\bf k}),{\bf H}({\bf k}) \right\}$ acquire the extra factor $\gamma^{-1}$, because the differential in the integrals (\ref{eq:Fourier}) is transformed as $d^3{\bf k}^\prime = \gamma\, d^3{\bf k}$ due to the Lorentz contraction. Considering the boost with the velocity ${\bf v} = v\, \bar{\bf x}$, this yields:
%
\begin{eqnarray}
{ {E}}^{'}_x = \gamma^{-1} { {E}}_x, & \quad & { {H}}^{'}_x = \gamma^{-1} { {H}}_x, \nonumber \\
{ {E}}^{'}_y =  { {E}}_y - u { {H}}_z , & \quad & { {H}}^{'}_y = { {H}}_y +u { {E}}_z , \nonumber \\
{ {E}}^{'}_z =  { {E}}_z + u { {H}}_y, & \quad & { {H}}^{'}_z = { {H}}_z - u { {E}}_y.    
\label{eq:fieldboost}
\end{eqnarray}  
%
This field transformation is accompanied by the Lorentz boost of the four-wavevector $(\omega/c,{\bf k})$:
%
\begin{eqnarray}
\omega^\prime = \gamma (\omega - v k_x), & \quad &
k_x^\prime = \gamma \left(k_x - \frac{v \omega}{c^2}\right), \nonumber \\
k_y^\prime = k_y,  & \quad & k_z^\prime = k_z.
\label{eq:kboost}
\end{eqnarray}  
%
The boosted fields in the helicity representation, $\left\{ \tilde{\bf E}^\prime({\bf k}^\prime), \tilde{\bf H}^\prime({\bf k}^\prime) \right\}$, are obtained from the fields (\ref{eq:fieldboost}) via the unitary transformation (\ref{eq:htransform}) involving the boosted wavevectors ${\bf k}^\prime$ (\ref{eq:kboost}) and the corresponding spherical angles $(\vartheta^\prime,\varphi^\prime)$.

\subsection{The Lorentz boost of a Bessel beam}

We are now in the position to consider a photon state carrying spin and orbital AM, the simplest model of which being provided by monochromatic {\it Bessel beams} \cite{Bliokh2010,Bliokh2017review,McGloin2005}. The Fourier spectrum of the $z$-propagating Bessel beam is a circle lying on the sphere of radius $k=k_0=\omega_0/c$ at the polar angle $\vartheta=\vartheta_0$, see Fig.~2(a). Assuming well-defined helicity $\sigma = \pm 1$ (i.e., the same right-hand or left-hand circular polarizations of all plane waves in the beam spectrum), the electric field of the Bessel beam can be written as \cite{Bliokh2010}:
%
\begin{equation} 
\label{eq:hBessel}
\tilde{{\bf E}}({\bf k}) = \frac{A}{2}\left( {\begin{array}{c}
   1 +\sigma  \\
   1 -\sigma  
\end{array}} \right) \delta(k-k_0) \delta(\vartheta-\vartheta_0) e^{i \ell\varphi},
\end{equation}
%
where $A$ is the field amplitude, $\delta$ is the delta-function, and $\exp(i \ell\phi) $ indicates a vortex with the integer topological charge $\ell$, which is responsible for the intrinsic orbital AM carried by the beam \cite{OAM,AML,Bliokh2015PR}. 

\begin{figure}[t]
\centering
\includegraphics[width=\columnwidth]{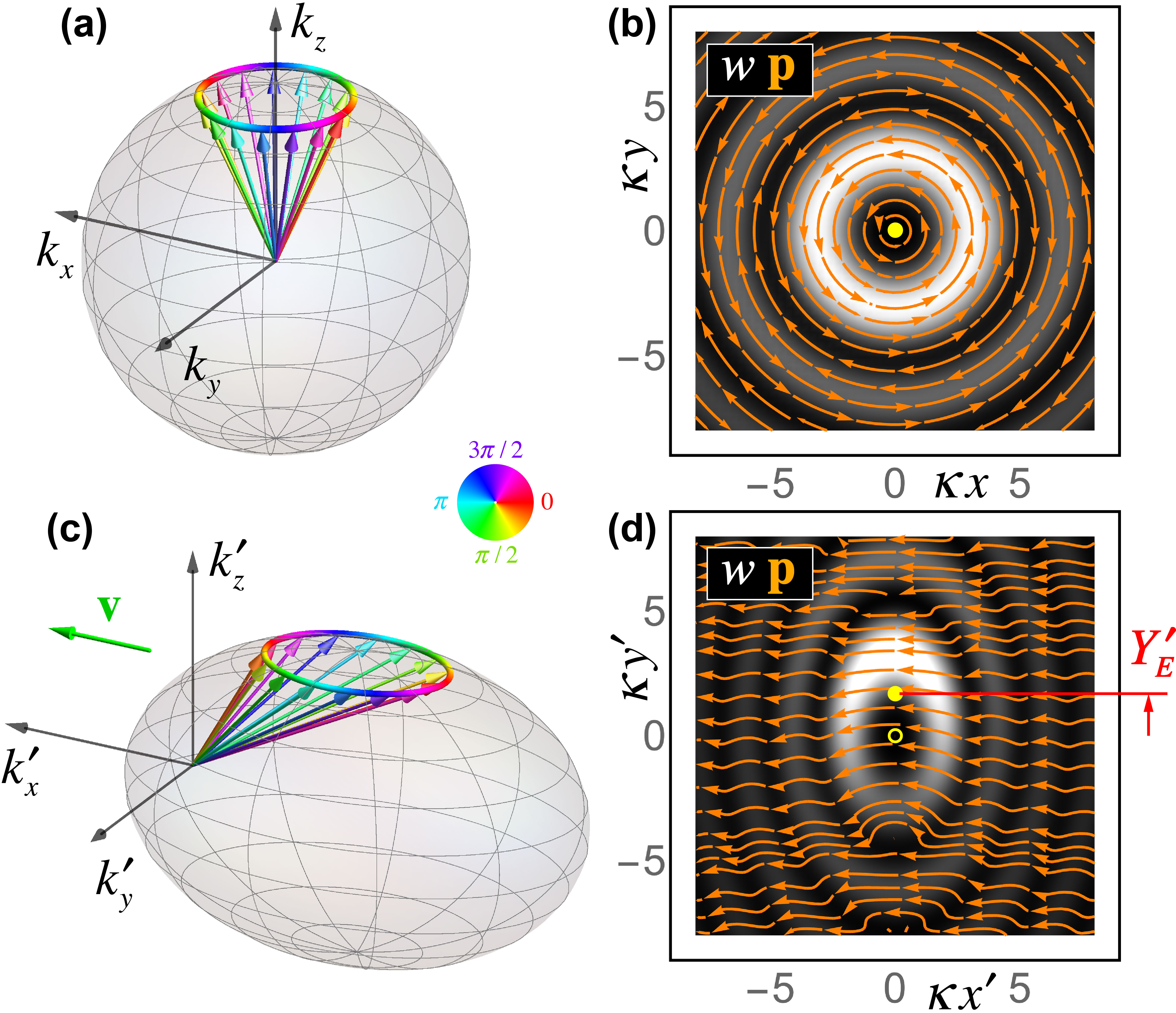}
\caption{Monochromatic $z$-propagating Bessel beam (\ref{eq:hBessel}) and (\ref{eq:Bessel}) (a,b) and the same beam in the reference frame moving with velocity ${\bf v}=v \bar{\bf x}$ (c,d). The Fourier spectra (i.e., the wavevector distributions with color-coded vortex phases $\exp(i\ell\varphi)$) (a,c) and the real-space distributions of the energy and Poynting-momentum densities (\ref{eq:densities}) (b,d) are shown.
One can see the non-monochromatic character ($\omega^\prime = k^\prime c \neq {\rm const}$) of the boosted beam, its elliptic Lorentz-contraction deformation, and the relativistic Hall-effect shift of the energy centroid: $Y_E^\prime = (v/\omega_0)(\ell+\sigma)$. For better visibility, we used nonparaxial beams with the following parameters: $\sigma =1$, $\ell=2$, $\sin\vartheta_0 = 0.4$ (a,c), $\sin\vartheta_0 = 0.7$ (b,d),  $u = 0.8$ (c,d), and $\kappa = k_0 \sin\vartheta_0$ (b,d).}
\label{Fig_2}
\end{figure}

Using Eq.~(\ref{eq:htransform}), we obtain the Bessel-beam field components in the Cartesian coordinates:
%
\begin{equation} 
\label{eq:Bessel}
{{\bf E}}({\bf k})\! =\! \frac{A}{\sqrt{2}}\left( {\begin{array}{c}
   a - b e^{2i\sigma\varphi} \\
   i\sigma\!\left(a + b e^{2i\sigma\varphi}\right)  \\
-2 \sqrt{ab}\, e^{i\sigma\varphi}
\end{array}} \right)\! \delta(k-k_0) \delta(\vartheta-\vartheta_0) e^{i \ell\varphi},
\end{equation}
%
where $a=\cos^2\!(\vartheta_0/2)$ and $b=\sin^2\!(\vartheta_0/2)$. Since we are dealing with the helicity eigenstate, $\hat{\sigma}\tilde{{\bf E}}=\sigma \tilde{{\bf E}}$, the corresponding magnetic field is ${{\bf H}}({\bf k}) = - i\sigma {{\bf E}}({\bf k})$.

Evaluating the Fourier integrals (\ref{eq:Fourier}), we find the real Bessel-beam fields ${\bm {\mathcal{E}}}(t,{\bf r})$ and ${\bm {\mathcal{H}}}(t,{\bf r})$, and plot the trasverse real-space distributions of the energy and Poynting-momentum densities (\ref{eq:densities}) in Fig.~2(b). Paraxiality implies $\vartheta_0 \ll 1$, but the Bessel beams are exact solutions of Maxwell's equations for any values of $\vartheta_0$.

Calculating the expectation values (\ref{eq:wf})--(\ref{eq:operators}) of the energy, momentum, spin and orbital AM, etc., for the Bessel-beam field (\ref{eq:hBessel}), we obtain \cite{Bliokh2010}:
%
\begin{eqnarray}
\label{eq:EV0}
& W = \hbar \omega_0, \quad 
{\bf P} = \hbar  k_0 \cos{\vartheta_0} \bar{\bf z} \simeq \hbar  k_0 \bar{\bf z}, \nonumber \\ 
& {\bf L} = \hbar [\ell + \sigma (1 - \cos{\vartheta_0})] \bar{\bf z} \simeq \hbar \ell \bar{\bf z}, ~~  
{\bf S} = \hbar \sigma \cos{\vartheta_0} \bar{\bf z} \simeq \hbar \sigma \bar{\bf z}, \nonumber \\
& {\bf J} = \hbar (\sigma +\ell) \bar{\bf z}, \quad {\bf R}_{\perp} = {\bf N}_{\perp} = {\bf 0},
\end{eqnarray} 
%
where we used the paraxial approximation $\vartheta_0 \ll 1$, and the subscript $\perp$ indicates the transverse $(x,y)$ components. Note that the Bessel beams are delocalized in the longitudinal $z$-direction and are not square-integrable in the transverse plane. Therefore all the integrals of squared fields and the normalization factor ${\mathcal N}$ diverge but their ratios (\ref{eq:EV0}) are finite \cite{Bliokh2010,Bliokh2012}. The longitudinal photon and energy centroid coordinates $Z$ and $Z_E$ are ill-defined in the beam, but if we were to consider a long $z$-localized wavepacket, we would approximately obtain $Z = Z_E = (c^2{\bf P}/W) t = ct$. This corresponds to the vanishing boost momentum $N_z = 0$. Equations~(\ref{eq:EV0}) present the expected picture of a paraxial photon carrying intrinsic spin ($\sigma$) and orbital ($\ell$) AM \cite{Bliokh2015PR}.

We now perform the Lorentz boost (\ref{eq:kboost}) and (\ref{eq:fieldboost}) of the Bessel beam (\ref{eq:Bessel}). This brings about cumbersome but exact expressions for the boosted Bessel-beam fields $\left\{{\bf E}^\prime({\bf k}^\prime), {\bf H}^\prime({\bf k}^\prime) \right\}$, $\left\{ {\bm {\mathcal{E}}^\prime}(t^\prime, {\bf r}^\prime),{\bm {\mathcal{H}}}^\prime(t^\prime, {\bf r}^\prime) \right\}$, and the corresponding helicity-representation field $\tilde{\bf E}^\prime({\bf k}^\prime)$. Figures 2(c,d) show the Fourier spectrum and the real-space transverse distributions of the energy and Poynting momentum densities (\ref{eq:densities}) for these fields (cf. \cite{Bliokh2012,Stone2015,Bliokh2012PRA}).
One can see that the boosted field is not monochromatic anymore ($\omega^\prime = k^\prime c \neq {\rm const}$), it is elliptically deformed due to the Lorentz contraction, and its energy centroid is shifted in the transverse direction, $Y_E^\prime \neq 0$, which is a manifestation of the relativistic Hall effect \cite{Bliokh2012,Chen2014,Stone2015,Muller1992}. At the same time, since the helicity is Lorentz-invariant, the boosted field is still the helicity eigenstate: $\hat{\sigma}\tilde{{\bf E}}^\prime({\bf k}^\prime) = \sigma \tilde{{\bf E}}^\prime({\bf k}^\prime)$, $\sigma = \pm 1$, and ${{\bf H}}^\prime({\bf k}^\prime) = - i\sigma {{\bf E}^\prime}({\bf k}^\prime)$.

Most importantly, we can now calculate the expectation values (\ref{eq:wf})--(\ref{eq:operators}) for the boosted Bessel beam. In the $\vartheta_0 \ll 1$ approximation, this yields:
%
\begin{eqnarray}
\label{eq:EV1}
&& W^\prime = \hbar \gamma\, \omega_0~, \quad 
{\bf P}^\prime = \hbar  k_0\! \left(\bar{\bf z}^\prime - \gamma u \bar{\bf x}^\prime \right),  \nonumber \\ 
&& \color{blue}\boxed{\color{black} {\bf L}^\prime = \hbar [\ell\gamma + 
\color{NEW}\sigma(\gamma-\gamma^{-1}) \color{black}] \bar{\bf z}^\prime
+ \color{NEW} \hbar \sigma u\, \color{black} \bar{\bf x}^\prime}\color{black}~ ,  
 \nonumber \\
&& \color{blue}\boxed{\color{black} {\bf S}^\prime = \hbar \sigma\! \left[\color{black} \gamma^{-1} \color{black} \bar{\bf z}^\prime - \color{black} u\, \color{black} \bar{\bf x}^\prime\right]}\color{black}~, ~~ 
\color{NEW}\boxed{\color{black} {\bf J}^\prime = \hbar \gamma \left(\ell + \sigma \right) \bar{\bf z}^\prime } \color{black}~,  \\
&& \color{blue}\boxed{\color{black} {\bf R}_{\perp}^\prime\! =\! \frac{v}{2\omega_0} \left(\ell + \color{NEW} 2\sigma \color{black} \right)\! \bar{\bf y}^\prime - v t\, \bar{\bf x}^\prime}\color{black}~,~~ 
\color{NEW}\boxed{\color{black} {\bf N}_{\perp}^\prime\! =\! -\hbar\gamma u \left(\ell+\sigma \right) \bar{\bf y}^\prime}\color{black}~.\nonumber 
\end{eqnarray} 
%
These equations contain the central results of this work, which are also illustrated in Fig.~3. The energy and momentum (\ref{eq:EV1}) present the standard Lorentz transformation of the quantities (\ref{eq:EV0}):
$W^\prime = \gamma\, W$, ${\bf P}^\prime = {\bf P} - \gamma\, W {\bf v}/c^2$. The boost momentum also agrees with the Lorentz transformation of the relativistic AM tensor \cite{LL,Bliokh2012}: ${\bf N}^\prime = - \gamma\, {\bf J} \times {\bf v}/c$. This corresponds to the transverse Hall-effect shift of the energy centroid ${\bf R}_{E\perp}^\prime + {\bf v} t =  {\bf J} \times {\bf v}/W = (v/\omega_0)(\ell + \sigma) \bar{\bf y}^\prime $ [Fig.~2(d)], in agreement with recent results \cite{Bliokh2012} (for $\sigma=0$) and \cite{Chen2014,Duval2015,Stone2015} (for $\ell=0$).

At the same time, the AM parts and the ``photon centroid'' in Eqs.~(\ref{eq:EV1}) exhibit several unusual features. First, the spin AM is indeed transformed as expected for a polarized plane wave: ${\bf S}^\prime = \hbar\sigma {\bf P}^\prime/P^\prime$, Fig.~1(a), and in contrast to the relativistic AM transformation, Fig.~1(b). Second, this paradox is resolved by the nontrivial transformation of the orbital AM ${\bf L}^\prime$, which acquires unexpected {\it helicity-dependent terms}, both longitudinal and transverse. This signals the {\it relativistic spin-orbit interactions of light}, cf. \cite{Bliokh2015}. As a result, the total AM ${\bf J}^\prime = {\bf L}^\prime + {\bf S}^\prime$ is transformed exactly as expected for the relativistic AM with ${\bf N}= {\bf 0}$: ${\bf J}^\prime =  \gamma\, {\bf J}$.
Third, the photon centroid ${\bf R}_{\perp}^\prime$ exhibits the natural drift $-{\bf v}t$ in the moving frame and the {\it transverse Hall-effect shift} $Y^\prime = (v/2\omega_0)(\ell + 2\sigma)$. This differs from the previously analyzed spinless and massive-particle cases \cite{Muller1992,Bliokh2012} by the {\it factor of 2} before the helicity \cite{g-factor}.
This unexpected factor plays an important role in the Lorentz transformations (\ref{eq:EV1}) of the photon AM. 

\begin{figure}[t]
\centering
\includegraphics[width=0.85\columnwidth]{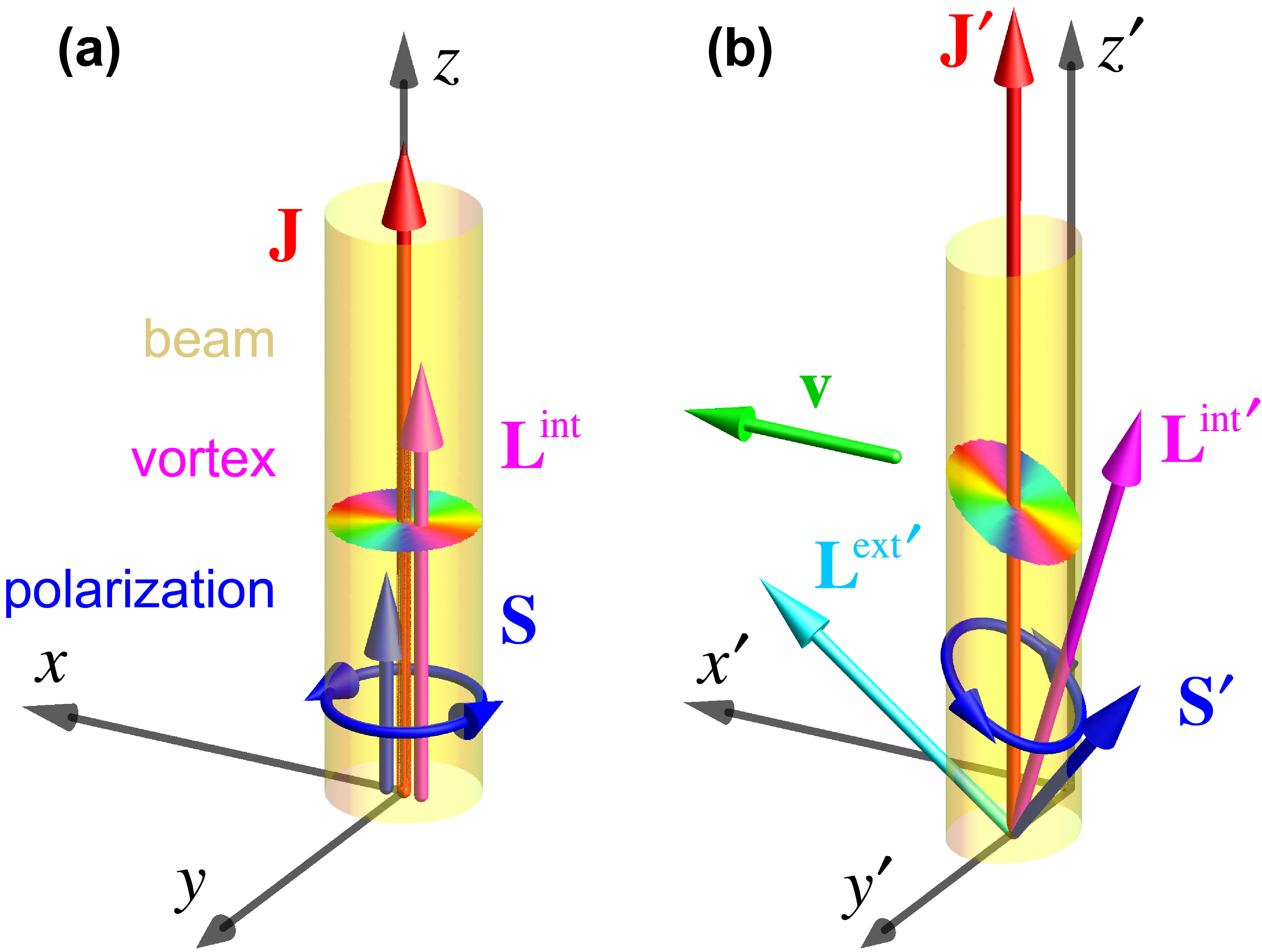}
\caption{Transformations of the spin and orbital AM in a paraxial vortex beam under a transverse Lorentz boost ($u = 0.6$ here). (a) The original monochromatic beam carries the spin AM ${\bf S}=\sigma {\bf P}/P$ due to the circular polarization (helicity) ($\sigma=1$ here), as well as the intrinsic orbital AM  ${\bf L}^{\rm int}=\ell\, {\bf P}/P$ due to the vortex ($\ell=2$ here). (b) The boosted beam carries spin AM
tilted together with the beam momentum: ${\bf S}^\prime=\sigma {\bf P}^\prime/P^\prime$ [Fig.~1(a)], the intrinsic orbital AM ${\bf L}^{{\rm int}\,\prime}$, Eq.~(\ref{eq:intrinsic}), due to the elliptically deformed and tilted vortex,
and the extrinsic orbital AM caused by the transverse shift (Hall effect) of the beam centroid: ${\bf L}^{{\rm ext}\, \prime} = {\bf R}^\prime \times {\bf P}^\prime$. Although all these contributions point in different directions, the total AM is transformed according to the Lorentz transformation: ${\bf J}^\prime = \gamma\, {\bf J}$ [Fig.~1(b)].
}
\label{Fig_3}
\end{figure}

Indeed, the photon centroid ${\bf R}^\prime$ allows us to separate the {\it intrinsic} (vortex-related) and {\it extrinsic} (shift-induced) contributions to the orbital AM \cite{Bliokh2015,Bliokh2010,Bliokh2012,Bliokh2015PR}:
%
\begin{eqnarray}
\label{eq:intrinsic}
{\bf L}^{{\rm ext}\,\prime} & = & {\bf R}^{\prime} \times {\bf P}^{\prime} = \hbar\left(\ell+ \color{NEW}2\sigma\color{black} \right)\left( \frac{\gamma-\gamma^{-1}}{2}\, \bar{\bf z}^{\prime} + \frac{u}{2}\, \bar{\bf x}^{\prime} \right),
\nonumber\\
{\bf L}^{{\rm int}\, \prime} & = & {\bf L}^{\prime} - {\bf L}^{{\rm ext}\, \prime} = \hbar\ell \left( \frac{\gamma+\gamma^{-1}}{2}\, \bar{\bf z}^{\prime}
- \color{NEW} \frac{u}{2} \, \bar{\bf x}^{\prime} \color{black} \right).
\end{eqnarray} 
%
Here we used the longitudinal photon position $Z^\prime =\gamma^{-1}ct$ because of the oblique propagation at the angle $\theta = \cos^{-1}\!\gamma^{-1}$. Remarkably, the form of the intrinsic orbital AM ${\bf L}^{{\rm int}\, \prime}$ can be clearly explained by the geometric deformations of the vortex phase in the beam. Namely, the vortex is {\it elliptically deformed} due to the Lorentz contraction with the factor of $\gamma$ and also {\it tilted} by the angle $\theta$, as shown in Fig.~3(b) (because the phase fronts in the boosted beam are near-perpendicular to the momentum ${\bf P}^\prime$). It is easy to show that these deformations, $x \to x/\gamma$, $k_x \to \gamma k_x$, and $z = x \tan \theta  = u\gamma x \to ux$, result in the intrinsic orbital AM (\ref{eq:intrinsic}) for a vortex wavefunction $\psi \propto (x+iy)^\ell$. Importantly, the $x^\prime$-directed term in ${\bf L}^{{\rm int}\, \prime}$, related to the tilt of the vortex, was missed in previous studies \cite{Bliokh2012,Bliokh2012PRA} only focused on the longitudinal $z^\prime$-component of the AM. The set of equations (\ref{eq:EV1}) and (\ref{eq:intrinsic}) show that both this new term and the factor of 2 before the helicity in the centroid shift ${\bf R}^\prime$ ensure the proper Lorentz transformation of the total AM ${\bf J}$.

In addition to the analytical ${\bf k}$-space calculations of the expectation values (\ref{eq:EV1}), we numerical calculated the values $W^\prime$, ${\bf P}^\prime$, ${\bf J}^\prime$, and ${\bf N}^\prime$ using the ${\bf r}$-space integration of the energy and Poynting-momentum densities (\ref{eq:densities}) in the transformed Bessel beam. The results were in agreement with Eqs.~(\ref{eq:EV1}). Here we should make two important remarks. First, since Bessel beams are delocalized along the longitudinal $z$-axis, the integration should be performed over a 2D cross-section of the beam. In doing so, the result depends on the choice of the cross-section, similar to the ``geometric spin Hall effect of light'' \cite{Aiello2009,Korger2014}. We found that the proper Lorentz transformation of the AM is obtained using the integration in the {\it tilted plane} $z^\prime =u\gamma x^\prime$ parallel to the phase fronts (i.e., orthogonal to the momentum) of the boosted beam, Fig.~3(b). This is in agreement with the Wigner-translation approach used in \cite{Stone2015}. In the ${\bf k}$-space calculations of Eqs.~(\ref{eq:EV1}) the tilted-cross-section condition was also used as $\partial/\partial k_z^\prime = u\gamma\, \partial/\partial k_x^\prime$. Second, we note that the Berry connection $\hat{\bf A}_B ({\bf k}^\prime)$ in the operators (\ref{eq:operators}) played a crucial role in obtaining the transformed quantities (\ref{eq:EV1}). In the paraxial limit $\vartheta_0 \to 0$, it is determined by the mean momentum ${\bf P}^\prime$ and equals $\hat{\bf A}_B = - \hat{\sigma} k_0^{-1} [(1-\gamma^{-1})/(\gamma u)]\, \bar{\bf y}$. This illuminates the geometric SOI origin of the nontrivial transformations (\ref{eq:EV1}) \cite{Bliokh2015,Bliokh2010}.

\section{Relativistic transformations of Dirac-electron beams}
\subsection{General formalism}
 
It is interesting to check if the nontrivial transformations (\ref{eq:EV1}) and (\ref{eq:intrinsic}) of the spin and orbital AM quantities are specific to photons (i.e., massless spin-1 particles) or these have a universal character. For this purpose, we consider a similar Lorentz-transformation problem for a Bessel-beam state of the {\it Dirac electron} \cite{Bliokh2011}, i.e., a massive spin-1/2 particle.

We first recall the Dirac equation in the standard representation \cite{QED}:
%
\begin{equation}
i \hbar \frac{\partial {\bm \psi}}{\partial t} = \left( \hat{\bm \alpha}\cdot \hat{\bf P}c + \hat{\beta}\, m c^2 \right) {\bm \psi}~,
\label{eq:DE}
\end{equation}
%
where ${\bm \psi}(\bf{r},t)$ is the four-component bi-spinor wavefunction, $\hat{\bf P} = -i \hbar \bm{\nabla}$ is the momentum operator in the coordinate representation, $m$ is the electron mass, and
%
\[  \hat{\bm \alpha} = \left(\begin{array}{cc} 0 & \hat{\bm \sigma} \\ \hat{\bm \sigma} & 0 \end{array}\right), \qquad
\hat{\beta} = \left(\begin{array}{cc} 1 & 0 \\ 0 & -1 \end{array}\right)  \]
%
are the $4\times 4$ Dirac matrices with $\hat{\bm \sigma}$ being the vector of the $2\times 2$ Pauli matrices. 

The wavefunction can be represented as the Fourier integral, i.e., as a superposition of Dirac plane waves:
%
\begin{equation}
{\bm \psi} ( {{\bf{r}},t} ) = \int {\frac{d^3 {\bf k}}{(2\pi)^{3/2}}} \,\tilde {\bm \psi}({\bf k})\,
e^{- i \omega t + i {\bf k} \cdot {\bf r}}~.
\label{eq:DEFourier}
\end{equation}
%
Here, $\omega ({\bf k}) = \sqrt{k^2 c^2 + \mu^2}$, $\mu = mc^2/\hbar$, and the Fourier amplitudes can be factorized as:
%
\begin{equation}
\tilde{\bm \psi}({\bf k})\! = \! f({\bf k}) {\bm \Phi}({\bf k}),~~
{\bm \Phi}({\bf k}) \!=\! \frac{1}{\sqrt {2\omega } }\left( {\begin{array}{*{20}{c}}
{\sqrt{\omega  + \mu} \; {\bm \xi}}\\
{\sqrt {\omega  - \mu} \,\left( {\hat{\bm \sigma} \cdot {\bf{\bar k}}} \right) {\bm \xi}}
\end{array}} \right),
\label{eq:DEplane}
\end{equation}
%
where $f({\bf k})$ is the scalar Fourier amplitude, ${\bm \Phi}({\bf k})$ is the normalized polarization bi-spinor (${\bm \Phi}^{\dag} {\bm \Phi}=1$), ${\bf{\bar k}} = {\bf k}/k$, and ${\bm \xi} = \left( {\begin{array}{*{20}{c}} a\\ b \end{array}} \right)$
is the two-component polarization spinor (${\bm \xi}^{\dag} {\bm \xi}=1$) describing the spin state of the plane-wave electron in its rest frame \cite{QED,Bliokh2011,Bliokh2017review}.

The Fourier amplitudes $\tilde{\bm \psi}({\bf k})$ can be regarded as the (non-normalized) Dirac wavefunction in the canonical momentum representation. In this representation, the operators of the energy, momentum, position, spin, orbital angular momentum, and boost momentum have a canonical form similar to Eq.~(\ref{eq:operatorscan}):
%
\begin{eqnarray}
\label{eq:DEoperators}
{\hat{W}}=\hbar\omega, & \quad & {\hat{\bf P}}=\hbar{\bf k},  \quad {\hat{\bf R}}= i  {\bm{\nabla}}_{{\bf k}}, \\
{\hat{\bf S}}= \frac{\hbar}{2} \left(\begin{array}{cc} 0 & \hat{\bm \sigma} \\ \hat{\bm \sigma} & 0 \end{array}\right), & \quad &
{\hat{\bf L}}= {\hat{{\bf R}}} \times {\hat{\bf P}},  \quad  {\hat{\bf N}} = ct \hat{\bf P} - \hat{\bf R}\hat{W}/c .  \nonumber
\end{eqnarray}  
%
The normalized (one-electron) expectation values are calculated similar to Eq.~(\ref{eq:EV}):
%
\begin{equation}
O = \frac{1}{{\mathcal N}} \langle \tilde{\bm \psi} | \hat{O} | \tilde{\bm \psi} \rangle
= \frac{1}{{\mathcal N}} \int d^3{\bf k}\, \tilde{\bm \psi}^{\dag}({\bf k})\! \cdot \! (\hat{O}) \tilde{\bm \psi}({\bf k})~,
\label{eq:DEEV}
\end{equation}  
%
with the number of electrons ${\mathcal N} = \int d^3{\bf k}\, |\tilde{\bm \psi}({\bf k})|^2$. Note that, in contrast to photons, the inner product for electrons does not involve the $\omega^{-1}({\bf k})$ factor. This is because the squared wavefunction amplitudes correspond to the particle and energy densities for electrons and photons, respectively. 

It is worth remarking that one can alternatively use the Foldy-Wouthuysen momentum representation for the calculation of the expectation values for Dirac electrons. This representation, diagonalizing the Dirac Hamiltonian, allows one to reduce the wavefunction to the two components ${\bm \xi}({\bf k})$, but complicates the operators with the Berry-connection terms, similar to the helicity representation (\ref{eq:htransform}) and (\ref{eq:operators}) for photons \cite{Bliokh2011,Berard2006,Bliokh2005,Bliokh2017}.

We finally introduce the Lorentz transformation (with the velocity ${\bf v} = v\, \bar{\bf x}$) of the Dirac wavefunction $\tilde{\bm \psi}({\bf k})$. Akin to the transformation of the Fourier components of Maxwell fields, Eq.~(\ref{eq:fieldboost}), it acquires an extra $\gamma^{-1}$ factor and reads \cite{QED}:
%
\begin{equation}
\tilde{\bm \psi}' = \frac{1}{{\sqrt 2 \,\gamma }}\left( {\sqrt {\gamma  + 1}  - \sqrt {\gamma  - 1}\; \hat{\alpha}_x } \right) \tilde{\bm \psi}~.
\label{eq:DEwfboost}
\end{equation}  
%
The Lorentz transformation of the electron four-wavevector $(\omega/c,{\bf k})$ is still given by Eq.~(\ref{eq:kboost}).

\subsection{The Lorentz boost of a Dirac-Bessel beam}

The Bessel-beam state of the Dirac electron (i.e., the Dirac-Bessel beam) is constructed similar to the optical beam (\ref{eq:hBessel}) and (\ref{eq:Bessel}), Fig.~\ref{Fig_2}(a). The scalar part and the polarization spinors of the two spin states of the electron are given by \cite{Bliokh2011,Bliokh2017review}:
%
\begin{equation}
f({\bf k}) \!=\! A \delta(k-k_0) \delta(\theta-\theta_0) e^{i\ell\phi}, ~~
{\bm \xi}^+ \!=\! \left( {\begin{array}{*{20}{c}} 1 \\ 0 \end{array}} \right),~~
{\bm \xi}^- \!=\! \left( {\begin{array}{*{20}{c}} 0 \\ 1 \end{array}} \right).
\label{eq:DiracBessel}
\end{equation}  
%
These states correspond to the well-defined $z$-components of the electron spin, $s_z = \pm 1/2$, in its rest frame. Alternatively, one can choose two states with well-defined helicity \cite{Bliokh2017review,QED}, but these states reduce to the same ${\bm \xi}^\pm$ states in the paraxial approximation $\theta_0 \ll 1$. 

Substsituting the wavefunction (\ref{eq:DEplane}) and (\ref{eq:DiracBessel}) into Eqs.~(\ref{eq:DEoperators}) and (\ref{eq:DEEV}), we obtain the expectation values of the energy, momentum, AM, etc. for the Bessel-Dirac electron in the paraxial limit:
%
\begin{eqnarray}
\label{eq:DEEV0}
& W = \hbar \omega_0, \quad 
{\bf P} = \hbar  k_0 \bar{\bf z}, \quad 
{\bf R}_{\perp} = {\bf N}_{\perp} = {\bf 0}, \nonumber \\ 
& {\bf L} = \hbar \ell \bar{\bf z}, \quad  
{\bf S} = \hbar s_z \bar{\bf z}, \quad
 {\bf J} = \hbar (s_z +\ell) \bar{\bf z}. 
\end{eqnarray}  
%
This coinsides with Eqs.~(\ref{eq:EV0}) with the only difference that now $\omega_0 = \sqrt{k_0^2c^2 +\mu^2}$, and the spin quantum number $s_z=\pm 1/2$ substitutes the helicity $\sigma = \pm 1$. Note that the longitudinal boost momentum also vanishes, $N_z =0$, when we assume the relativistic equation of motion $Z=Z_E = (c^2 {\bf P}/W)\,t$.

Now, performing the Lorentz transformation (\ref{eq:DEwfboost}) and (\ref{eq:kboost}) of the Dirac-Bessel wavefunction (\ref{eq:DEplane}) and (\ref{eq:DiracBessel}), we calculate the expectation values (\ref{eq:DEoperators}) and (\ref{eq:DEEV}) in the boosted reference frame. Remarkably, this results in formulae very similar to photonic Eqs.~(\ref{eq:EV1}):
%
\begin{eqnarray}
\label{eq:DEEV1}
&& W^\prime = \hbar \gamma\, \omega_0~, \quad 
{\bf P}^\prime = \hbar  k_0\! \left(\bar{\bf z}^\prime - \gamma u \frac{\omega_0}{k_0c} \bar{\bf x}^\prime \right),  \nonumber \\ 
&& {\bf L}^\prime = \hbar [\ell\gamma + 
s_z (\gamma-\gamma^{-1})] \bar{\bf z}^\prime
+  \hbar s_z u \frac{k_0c}{\omega_0} \bar{\bf x}^\prime~ ,  
 \nonumber \\
&& {\bf S}^\prime = \hbar s_z\! \left( \gamma^{-1}  \bar{\bf z}^\prime - u \frac{k_0c}{\omega_0} \bar{\bf x}^\prime\right), ~~ 
 {\bf J}^\prime = \hbar \gamma \left(\ell + s_z \right) \bar{\bf z}^\prime ~,  \\
&& {\bf R}_{\perp}^\prime\! =\! \frac{v}{2\omega_0} \left(\ell + 2s_z \right)\! \bar{\bf y}^\prime - v t\, \bar{\bf x}^\prime~,~~ 
{\bf N}_{\perp}^\prime\! =\! -\hbar\gamma u \left(\ell+s_z \right) \bar{\bf y}^\prime ~.\nonumber 
\end{eqnarray} 
%
The main difference from the photonic case is that $\omega_0 / k_0 c \neq 1$, and these factors modify the directions of the boosted momentum, spin, and orbital AM. 

One can see that the modified transformations (\ref{eq:DEEV1}) exactly correspond to the fact that the {\it Pauli-Lubanski four-pseudovector} $(\Sigma_0,{\bm \Sigma})=({\bf S}\!\cdot\!{\bf P},\,{\bf S}W/c)$ \cite{QED}, orthogonal to the electron four-momentum $(W/c,\,{\bf P})$, is transformed as a four-vector under Lorentz boosts. This has several consequences. First, the direction of the spin AM does not  follow the momentum of the electron: ${\bf S}' \nparallel {\bf P}'$. Second, the absolute value of the spin AM diminishes: $S^\prime = (\hbar/2)\sqrt{1-u^2\mu^2/\omega_0^2} < S$, which can be interpreted as partial {\it depolarization} of the boosted electron. Finally, the transformation of the Pauli-Lubaski vector also describes the transformation of the electron {\it helicity} $\hbar\sigma = \Sigma_0 /P$. In the original and boosted frames, the helicity becomes:
%
\begin{equation}
\label{eq:DEhelicity}
\sigma = s_z~,\quad
\sigma' = \frac{s_z}{\sqrt{1+u^2\frac{\mu^2}{k_0^2 c^2}}}~, 
\end{equation}  
%
which clearly indicates that the helicity is Lorentz-invariant only for massless particles.

Akin to the photonic case, Eqs.~(\ref{eq:intrinsic}), we separate the intrinsic (vortex-related) and extrinsic (shift-related) parts of the electron orbital AM:
%
\begin{eqnarray}
\label{eq:DEintrinsic}
{\bf L}^{{\rm ext}\,\prime} & = & {\bf R}^{\prime} \times {\bf P}^{\prime} = \hbar\left(\ell+ 2s_z \right)\left( \frac{\gamma-\gamma^{-1}}{2}\, \bar{\bf z}^{\prime} + \frac{u}{2}\frac{k_0c}{\omega_0}\, \bar{\bf x}^{\prime} \right),
\nonumber\\
{\bf L}^{{\rm int}\, \prime} & = & {\bf L}^{\prime} - {\bf L}^{{\rm ext}\, \prime} = \hbar\ell \left( \frac{\gamma+\gamma^{-1}}{2}\, \bar{\bf z}^{\prime}
- \ \frac{u}{2}\frac{k_0c}{\omega_0} \, \bar{\bf x}^{\prime} \right),
\end{eqnarray} 
%
where we used $Z'=(c^2P_z^\prime /W^\prime)\,t$. Thus, the intrinsic and extrinsic orbital AM of electrons are also analogous to those of photons (up to modification by the $k_0c/\omega_0$ factors). In particular, as it should be, the vortex-related intrinsic orbital AM depends only on the vortex quantum number $\ell$ and is independent of the spin $s_z$.

Similar to the case of optical beams, the expectation values of the $\hat{\bf R}$-dependent operators for electrons depend on the choice of the beam cross-section used in the integration. We found that Eqs.~(\ref{eq:DEEV1}), consistent with the Lorentz transformations of the relativistic AM, are obtained only when choosing the tilted cross section $z' = x' \tan \theta_S$, i.e. $\partial / \partial k_z^\prime = \tan \theta_S\, \partial/\partial k_x^\prime$, where $\theta_S = \tan^{-1}(u\gamma\, k_0c/\omega_0)$. Notably, this angle corresponds to the direction of the electron spin AM ${\bf S}'$ rather than momentum ${\bf P}'$ (for photons these directions coincide). Understanding this peculiarity requires further investigations of properly 3D localized Dirac wavepackets, which is beyond the scope of this study.

\section{Discussion}

We have considered relativistic transformations of the spin and orbital AM of paraxial photons and Dirac electrons under the transverse Lorentz boost. The main results are summarized in Eqs.~(\ref{eq:EV1}), (\ref{eq:intrinsic}), (\ref{eq:DEEV1}), (\ref{eq:DEintrinsic}), and in Fig.~3. We have found that the Lorentz transformations of these quantities, as well as of other beam characteristics, exhibit quite nontrivial forms, which together ensure the proper Lorentz transformation of the total AM and resolve the paradox with the transformation of the photon spin, Fig.~1. Most importantly, the transverse Lorentz boost inevitably produces the spin-to-orbital AM conversion (i.e., helicity-depend terms in the orbital AM) and nontrivial redistribution between the intrinsic (vortex) and extrinsic (shift) parts of the orbital AM. These effects have the geometric origin and evidence the {\it relativistic SOIs of light}.

Although we considered the particular case of Bessel beams (allowing analytical calculations), the results are generic for paraxial azimuthally-symmetric beams or wavepackets. This is because all derived transformations have very clear geometric/relativistic explanations, independent of the particular type of the beam. Note also that we considered only {\it transverse} Lorentz boosts. It is easy to see that a {\it longitudinal} $z$-boost does not break the monchromaticity of the beam and can only modify its parameters (\ref{eq:EV0}). Until this breaks the paraxiality of the beam (i.e., for $\gamma \ll \vartheta_0^{-1}$), the spin and orbital AM (\ref{eq:EV0}) remain practically unchanged. We also note that the general formalism developed in this work alows one to perform the Lorentz transformations of {\it arbitrary} Maxwell and Dirac fields and to determine their properties in any reference frame. These results can play an important role in scattering processes involving relativistic particles carrying intrinsic AM, as well as in studies of light emitted by fast-moving bodies.

It should be emphasized that the nontrivial transformations found in this work are actually fixed by fundamental reasons. Namely, the transformation of the total AM ${\bf J}$ and the boost momentum ${\bf N}$ are determined by the Lorentz boost of the {\it rank-2 AM tensor}, while the spin AM ${\bf S}$ follows the boost of the {\it Pauli-Lubanski four-vector} (this is applied to both electrons and photons \cite{Ivan,Bliokh2017}). Hence, the total AM and spin AM inevitably obey different transformations. The difference between these two determines the nontrivial form of the orbital AM ${\bf L}$. Moreover, the orbital AM can be split into the extrinsic part ${\bf L}^{\rm ext}$ (determined by the position of the particle) and the intrinsic one ${\bf L}^{\rm int}$ (related to the vortex phase structure of the wavefunction). If we adopt the fact that the intrinsic contribution must depend only on the vortex quantum number $\ell$ (but not on the spin state $\sigma$ or $s_z$), this unambiguously determines the position shift ${\bf R}'$ proportional to $(\ell + 2\sigma)$ or $(\ell + 2s_z)$. Interestingly, such dependence was previously known only for the magnetic moment of the Dirac electron \cite{QED,Bliokh2011,Barut1981}, and was directly associated with the $g=2$ gyromagnetic factor for the electron spin. Our calculations show that this combination is universal for the relativistic Hall effect, independently of the spin and mass of the particle. 

The difference between relativistic transformations of the spin and orbital AM can also be compared with the difference in the commutation relations of the quantum-mechanical versions of these quantities \cite{Bliokh2010,Nienhuis,Bliokh2017}. Neither spin nor orbital AM operators (assuming their second-quantization or covariant Berry-connection forms) obey the canonical SO(3) commutation rules, while the total AM does. In a similar manner, neither spin nor orbital AM obeys the proper Lorentz transformation of the total AM. This nicely illuminates the intimate links between quantum and relativistic features inherent in the Maxwell and Dirac equations.

\section*{Acknowledgements}\label{sec:Acknowledgements}
This work was supported by the RIKEN iTHES Project, MURI Center for Dynamic Magneto-Optics via the AFOSR Award No. FA9550-14-1-0040, the Japan Society for the Promotion of Science (KAKENHI), the IMPACT program of JST, CREST grant No. JPMJCR1676, the Sir John Templeton Foundation, the RIKEN-AIST Challenge Research Fund, and the Australian Research Council.

\end{document}